**Human Factors in the LastPass Breach**

Niroop Sugunaraj

School of Electrical Engineering and Computer Science (SEECS)

College of Engineering and Mines (CEM)

University of North Dakota

May 2024

## Abstract

This paper examines the complex nature of cyber attacks through an analysis of the LastPass breach. It argues for the integration of human-centric considerations into cybersecurity measures, focusing on mitigating factors such as goal-directed behavior, cognitive overload, human biases (e.g., optimism, anchoring), and risky behaviors. Findings from an analysis of this breach offers support to the perspective that addressing both the human and technical dimensions of cyber defense can significantly enhance the resilience of cyber systems against complex threats. This means maintaining a balanced approach while simultaneously simplifying user interactions, making users aware of biases, and discouraging risky practices are essential for preventing cyber incidents.

**Human Factors in the LastPass Breach**

**Introduction**

Human-engineered systems are primarily designed to follow a three-pronged approach: a system that focuses on users, built with a purpose, and/or resilient enough to avoid failures (Remington et al., 2012). As highlighted by Pfleeger and Caputo (Pfleeger and Caputo, 2012), a realistic approach to a sound cyber security system is one where architects or developers of system architectures should be cognizant of behavioral sciences as they design, develop, and use them to enable users to be at the center of implementing good security practices. For example, the relevance of behavioral sciences to security can be factored through aspects such as biases (e.g., confirmation biases that lead to noticing and implementing evidence that supports a previously-held opinion) and decision-making (e.g., taking actions that underestimate risks that are willingly taken versus those that are out of an agent's control) (Schneier, 2008). Alerts and alarms are also considered as critical aspects of cyber systems, designed primarily to inform future decisions and point personnel towards potential threats. Alerts refer to messages that inform a user or developer that divert his/her attention to a developing situation (i.e., less urgent) whereas alarms are messages that warrant immediate action (i.e., more urgent) from its agent (Remington et al., 2012).

For this case study, we will examine the LastPass cyber breach (disclosed in December 2022) from a human-systems perspective, and offer potential ways human factors can be applied to mitigate or even prevent similar events. This study adds to the findings made by the author in a previous research on handling system failures (Sugunaraj, 2024). Particularly, this study considers the role of biases, keeping security best practices, and alarm types for better response systems.

LastPass, a popular password management service, was compromised as part of a sophisticated cyber attack in 2022 (BCS, 2023) - specifically between August and October - resulting in unauthorized access to user (i.e., employee) and customer data. Stated officially by LastPass itself (Toubba, 2023), there were two security incidents that led to this compromise - (1) a software engineer's corporate machine was hacked to steal sensitive information (e.g., source code); and (2) malware was delivered to a senior engineer's machine to exploit vulnerable third-party software. The malware (specifically a keylogger) allowed the attacker to gain access to the LastPass corporate vault after the engineer entered his master password and completed the multi-factor authentication step. While the technical specifics of the attack are beyond the scope of this study, this incident can be approached from the perspective of humans interacting with technology.

**Development**

The LastPass attack was a complex multi-faceted intrusion that exploited both technical vulnerabilities and human oversight(s) where multiple layers of security were bypassed, indicating a certain degree of technical expertise but also an understanding of human factors. Further analysis of the incident reveals that the following factors may have played a significant role in the breach:

1. **Established security practices**: In addition to using strong/unique master passwords (i.e., secure authentication) and minimizing the cognitive efforts needed to work with these passwords (e.g., through password managers), careful consideration should be given to third-party security. For example, the third-parties involved in this incident were Amazon and Plex (Toubba, 2023) (a media platform) though the latter has not explicitly confirmed the existence of the exploited vulnerabilities on their platform. While Amazon is well established in the technology space, it will have security principles that will vary from the contracting organization (i.e., LastPass). This can mean that employees in LastPass may need to interact with unnecessarily complex interfaces to monitor user activity and/or modify or override security notifications, and that security practices at these third-parties may be less stringent.

2. **Bias and systematic processing**: As far as the public is aware, there haven't been any major security incidents at LastPass; therefore, because of not having encountered any such incidents, the engineers may have developed a bias towards categorizing any potential threats or indiactors of compromise (IoCs) as non-issues. This can be considered as the interplay of the following human biases: (a) availability/heuristic bias - likelihood of events is judged on their ease of recall i.e., one can overestimate 'familiar' threats (even if they're not frequent) and underestimate more dangerous threats (i.e., potentially more frequent); and (b) anchoring bias - relying on the first piece of information more than necessary to make subsequent decisions. Additionally, there is an indication that the engineers may not have systematically processed the IoCs.

3. **Lack of a timely response system**: As mentioned officially by LastPass (Toubba, 2023), alerting and logging was enabled but logs did not immediately indicate anomalous behavior. It could be that alerts/alarms were not optimally implemented because of how unlikely it may have seemed for this breach to happen. The latter hasn't been confirmed by LastPass but is a reasonable assumption given how high false alarm rates can be (Remington et al., 2012).

## Analysis

This section will critically evaluate each factor presented in the previous section and jot down any limitations or caveats that may exist. While the factors contributing to the breach are multiple and an exhaustive study is required to analyze these factors, this study is constrained to what was deemed as the most obvious security aspects behind the compromise that can be looked at from a human factors perspective. It should also be mentioned that the mitigation measures under these factors can be considered "hypothetical" as sufficient details behind this attack are not available. However, efforts are taken to empirically correlate existing details about this attack to human factors.

**Security Practices**

Firstly, popular third-parties such as Paypal, Stripe, Venmo, Amazon etc. all have various design principles that will vary from their contracting organization(s) (e.g., LastPass). This leads to security teams navigating less user-friendly interfaces for customer behaviours/system overviews among other key tasks. Secondly, security practices at these third-parties may be less stringent depending on their maturity in the cyber space; this can contribute to misconfigured applications that LastPass' employees may not be aware of, sub-optimal employee password practices by third-parties, or lack of fail-safe mechanisms in case of a cyber incident. More specifically, complex interfaces contribute to features and frameworks that not only add to the cognitive workload for professionals, but also contain components and libraries that support otherwise avoidable attack vectors therefore making them more susceptible to attacks. Similarly, unprotected storage or storing passwords in recoverable formats (i.e., ciphertexts) create risks such as user access, brute force attacks, and privilege escalation (Crashtest Security, n.d.).

From LastPass' perspective, baseline security standards can be specified by its security team that third-party providers must comply with (US Department of Treasury, 2022). Baseline security standards such as tracking hardware and supply chains, identifying and updating single points of failure, and adequate information-sharing between third-parties enhances the security for the LastPass and its third-parties. For example, information-sharing in particular can enhance all three levels of situation awareness by keeping only relevant elements in the environment, improved comprehension through expert knowledge transfers, and project future actions based on the experience of operators in different third parties. Also, security training programmes are encouraged to be designed with the intention of informing users (i.e., employees) about any monetary penalties or other sanctions in case users do not comply with security best practices or voluntarily engage in risky security behaviors. Users who are clearly informed of certainty of consequences of violating security policies are more likely to be secure in their interactions (Chowdhury et al., 2020). Anchoring programmes with this tactic have been shown to be antecedents to successful policy compliance (Tsohou et al., 2015). According to Kennison and Chan-Tin (Kennison and Chan-Tin, 2020), creating "victim profiles" i.e., profiles that track variables such as general risk taking in daily life and the Big Five characteristics (Openness, Conscientiousness, Extraversion, Agreeableness, and Neuroticism) of such users can aid in organization in targeting and improving security training. Such an approach might offer the benefit that organizations would have done what is feasible and in their power to inform individuals of security best practices. However, there might be obvious privacy concerns with this approach because users may not be willing to share their information and may choose to opt out of any surveys or other forms that gather such data.

**Bias and Systematic Processing**

The research that is quite relevant - but not limited - to this factor would be those that consider goal-directed behaviour and human biases. For example, victims who have their accounts compromised would benefit greatly from assessing a potential cyber threat or cyber situation through their cognitive lens and less through their perception-action reflex, especially when non-critical time constraints are involved. In the case of the LastPass hack, simply being more deliberate in taking certain actions (e.g., checking for any unusual user activity) with the primary goal of keeping oneself secure from a cyber perspective could have played a major role in detecting the keylogger. While hindight is always 20/20, there might have been a sense of urgency for the developer in finishing certain tasks, possibly hinting at increased probability of neglecting the goal (i.e., being secure while completing his tasks) (Duncan et al., 1996). Additionally, the employees may have likely been victims to the anchoring bias i.e., dismissing any IoCs on seemingly reassuring surface-level cues (e.g., lack of alerts, sluggish CPU performance) and using these cues as an anchor to continue business as usual. As mentioned by Bahreini and colleagues (Bahreini et al., 2023), users do not take time to make secure decisions (e.g., validating that credentials have not been accessed from another location) due to the inconveniences that said decisions may lead to. Their study lends support to the fact that users take the more convenient route when making security decisions; in this case, study participants used search engines and online advice (i.e., anchors), rather than going to experts in their organization for a more credible answer to the same question.

Optimism bias is defined as the tendency to overestimate positive outcomes and underestimate negative outcomes. Because of this, it's easy to ignore IoCs (e.g., a delay in keystrokes, slow/laggy applications), through the delicate and subliminal interactions between factors within structure and content of the system (Remington et al., 2012). In this case, for example, it's possible that having used his/her home computer many times without falling victim to a hack, the bias would have led the developer to believe that nothing is amiss with his computer despite a keylogger running in the background which can ordinarily be tracked by noticing CPU usages, deploying an antivirus, or noticing relevant IoCs. Similarly, the factors of structure would have led the developer to assume that most of the elements in his interaction (e.g., no computer crashes or notifications indicating a breach) require no scrutiny simply because it is something one has seen before a good number of times. Adding to this is the point connecting cognition and goals; one is more motivated to "see" things that one is actively looking for and may overlook other things simply because those things are not part of one's objective(s).

Systematically processing information would involve a deeper analysis by looking at CPU usages, notifications of compromised credentials, evidence of uninitiated actions from a legitimate user, etc. This can be understood better by considering the heuristic-systematic theory. A study by Chaiken and Ledgerwood (Chaiken and Ledgerwood, 2012) states that those who retain information better (e.g., relayed through security programmmes) with low ego-involvement (and therefore higher systematic processing) ignored surface-level details such as the way information is delivered (e.g., tone) and rather focused on the logic/argumentation for an issue. Recalling and maintaining information that is produced in this way (i.e., from systematic processing) enables healthier security. Thus, decision-making theory is another way to bring awareness to people on how to interact better (i.e., more securely) with information systems by improving the way information is presented, retained, and retrieved.

**Response Systems**

From the perspective of response systems (i.e., alerts and alarms), research for auditory and aural alarm modalities by Bliss et al. (Bliss et al., 1995) shows that in cases where systems trigger false alarms quite frequently (e.g., misconfigurations/poor policies) leading to high false alarm rates, there is a degradation of response in terms of increased physiological stress and predicted response behavior. Additionally, looking further at research in time-critical fields (e.g., healthcare, aviation) will help in applying principles to the security domain. For example, using multi-modal alarms during medical procedures where practitioners are alerted to changes in a patient's vital signs through auditory alarms that are complemented by visual displays (Manish and RAJ, 2009) have shown to be successful. In the LastPass case, using non-visual alarm types such as auditory (e.g., beeps) or vibratory (e.g., haptic feedback) signals and different visual artifacts could have notified the victim before it escalated. Research supporting this way of presenting alarms is done by Vance and colleagues in 2018 (Vance et al., 2018) where "polymorphic" warnings were used to quantify user attention. Polymorphic warnings are those with unique artifacts such as using a background color of red or using a zoom animation to increase the size of the warning. Findings showed that such warnings maintained higher levels of attention and adherence compared to static warnings. Referring back to the section on security practices, it will help to inform users not to override or disable any alerts that come set according to corporate policy and to inform them of the consequences of not following recommendations, as it is not uncommon for users to disable alerts due to large false alarm rates (Manish and RAJ, 2009).

The proper use of alarms must make it immediately distinguishable from other elements that the users is dealing with. For example, from a visual standpoint (e.g., information display), alarms such as security popups should be clearly distinct from other ordinary aspects of the interface and be placed in a location on the display where the gaze of the user easily falls (Krol et al., 2012). Additionally, users must be kept from being de-sensitized to alarms by reducing the frequency at which such warnings are generated. This again comes down to designing alarms intelligently to reduce false positive and false alarm rates.

## Conclusion

Cyber systems thrive on the interactions between the information system and the user (i.e., human elements). Therefore, cyber incidents or attacks atleast to a degree are a function of system aspects (e.g., design practices, policies) and human factors (i.e., memory, user friendliness). The LastPass cyber attack was primarily due to two incidents that gave elevated user privileges to the hacker to access LastPass' corporate vault that contained employee and customer credentials. Considering cognitive overloads (i.e., reducing unnecessarily complex usage that encourage risky behaviors), policy design (i.e., security training and consequences of violation), human factors such as goal-directed behavior (i.e., placing an emphasis on security above all else), bias (i.e., underestimating negative outcomes/overestimating positive outcomes, assigning anchors), and risky behaviors are considerations to combat cyber incidents like the LastPass breach.